# Mid-IR single- and dual-electro-optic comb generation with an ultrafast modulator

G.L Ngo[1,*], L. Lucia[1], S. Pirotta[1], M. Malerba[2], L.H. Li[4], S. Barbieri[3], V. Merupo[3], E.H. Linfield[4], B. Darquié[5], JM. Manceau[1], A. Bousseksou[1], R. Colombelli[1,#]

[1]*Centre de Nanosciences et de Nanotechnologies (C2N), CNRS UMR 9001, Université Paris-Saclay, Palaiseau 91120, France.*
[2]*Istituto Nazionale di Ricerca Metrologica (INRiM), 10135 Torino, Italy*
[3]*IEMN, Univ. Lille, ISEN, CNRS, UMR 8520, 59652 Villeneuve d'Ascq, France*
[4]*School of Electronic and Electrical Engineering, University of Leeds, Leeds LS2 9JT, United Kingdom*
[5]*Laboratoire de Physique des Lasers, CNRS UMR 7538, Université Sorbonne Paris Nord, Villetaneuse, France*

Abstract:

Mid-infrared (mid-IR) frequency combs are powerful tools for molecular sensing, industrial monitoring, and precision spectroscopy, yet their development beyond 5 µm remains limited. Electro-optic modulation offers a promising path toward compact, agile comb generation, but extending this approach into the mid-infrared has been hindered by the lack of practical, high-performance modulators. Here we present an approach that leads to efficient generation of mid infrared frequency combs around 9 µm, by employing ultrafast, room temperature, free space electro optic intensity modulators. By driving a single modulator with short electrical pulse trains, we realize both single- and dual-comb operation from a continuous-wave quantum cascade laser, providing a compact and versatile platform for mid-IR spectroscopy. This scheme produces combs with tunable repetition rates down to the megahertz range with direct observation on an electrical spectrum analyzer without any interferometer. As a proof of concept, we perform single- and dual-comb spectroscopy of a germanium etalon and an ammonia cell, achieving resolution far beyond that of conventional Fourier-transform infrared (FTIR) spectrometers and highlighting the potential of this approach for precise measurements in the long-wavelength molecular-fingerprint region. These results establish high-performance mid-IR modulators as a promising route toward practical, energy-efficient frequency-comb systems for sensing and spectroscopy.

* Email: gia-long.ngo@cnrs.fr
# Email: raffaele.colombelli@c2n.upsaclay.fr

## 1. Introduction

The invention and development of the frequency comb (FC) has opened a vast range of opportunities in across multiple application domains [1]. In the visible and near-infrared (VIS/NIR) spectral regions, this technology has enabled transformative advances in metrology and spectroscopy [2]. In the mid-IR, where molecular 'fingerprints' allow sensitive trace-gas detection [3], with major applications in security, industrial process monitoring, and environmental sensing, the availability of practical FC sources is particularly important. Among the various approaches, dual-comb spectroscopy (DCS) is especially powerful [4]: a key advantage is the ability to acquire spectra at much higher speeds than conventional FTIR spectrometers (hundreds of µs rather than tens of seconds) [5], [6], thanks to its unique capability to operate in a parallel spectral acquisition mode and its lack of moving parts.

Frequency combs are traditionally generated using mode-locked lasers, where a fixed phase relationship between the longitudinal cavity modes produces a train of (ultra)-short pulses in the time domain (TiSa, for instance) [7]. Consequently, a major current challenge in the field is the realization of mode-locking in *compact* mid-IR laser sources, and thus the generation of mid-IR FCs. At present, the most mature *compact* mid-IR FC technology (we exclude therefore OPO and OPA systems [8] [9]) relies on quantum cascade lasers (QCLs) operated under RF injection [10] [11]. Parallel efforts are aimed at developing FCs based on interband cascade [12] [13] and fiber lasers [14]. Dual comb spectroscopy has been demonstrated with those sources [15] [16].

An alternative route to (mid-IR) FC generation relies on ultra-fast amplitude or phase modulators, in a scheme known as electro-optic (EO) comb generation. In this approach, a train of short electrical pulses is applied to a modulator (MD), and transferred onto a continuous-wave (CW) laser carrier as an amplitude or phase modulation [17] [18]. Beyond simplifying the laser source, a simple CW laser EO comb generation offers exceptional flexibility in the repetition rate, which can be tuned almost arbitrarily. This feature is particularly valuable in the mid-IR, as the FC repetition rate – together with the laser linewidth - sets the maximum achievable spectral resolution in spectroscopic measurements. Whereas QCL-based combs typically exhibit repetition rates ($f_{rep}$) in 10-20 GHz range, EO combs can reach $f_{rep}$ of only a few MHz: this enables spectral resolutions that are far more suitable for trace-gas detection.

Although EO comb generation is established in the VIS/NIR [19], in the mid-IR such approach to FC generation has been hindered by the lack of efficient, ultra-fast modulators. The only commercially available mid-IR modulation devices are bulky germanium acousto-optic modulators (AOMs) with narrow (~1MHz) modulation bandwidths (BW), up to ~100MHz; or narrow band (~100kHz) resonant electro-optic (EO) phase MDs based on bulk GaAs and CdTe with frequencies up to ~20GHz [20] [21]. Recently, advances in semi-guided and guided architectures have been reported on III–V [22] [23] and on SiGe semiconductor platforms [24] [25]. However, owing to limited bandwidth or insufficient modulation contrast, efficient EO comb generation in the mid-IR has not yet been demonstrated [25].

We have pioneered free-space-operating, ultra-fast mid-IR modulators based on the ultra-fast switching between strong and weak light matter coupling [26], and recently demonstrated operation with 3dB cut-off at ~10 GHz [27], at wavelengths around 9.6 µm. Here, we significantly advance their performance to demonstrate – for the first time to our knowledge – EO frequency comb generation in the mid-IR and suggest a technological alternative for high-accuracy spectroscopic measurements. This work does not primarily focus on modulator performance; instead, it highlights the novel and potentially transformative developments enabled by these devices. Specifically, we demonstrate efficient generation of mid-infrared EO FCs around 9 µm, achieve EO dual-comb generation, and perform both single- and dual-comb spectroscopy of a germanium etalon and an ammonia cell. The resulting spectral resolution far exceeds that of conventional FTIR spectroscopy, underscoring the considerably vast potential of this approach.

## 2. Results

### 2.1. Ultrafast mid-IR intensity modulation and electro-optic comb principle

The ultra-fast mid-IR amplitude modulator used in this work consists in a more advanced generation compared to those previously reported in Ref. [27]. The operating principle remains unchanged, and the active region relies on a coupled asymmetric InGaAs/AlInAs quantum-well (QW) heterostructure [28] embedded in a metal–metal patch cavity. Modulation of the reflected beam arises from the ultra-fast modification of the surface reflectivity when an applied bias switches the system from the weak- to the strong-coupling regime. The active region is largely identical, except that the current-blocking $HfO_2$ layer has been replaced with a thickened InGaAs buffer, and the number of QWs has been increased. Furthermore, significant improvements were introduced to the fabrication process (see Methods) by transitioning from epoxy bonding to thermo-compressive wafer bonding, that provides substantially enhanced thermal stability and reliability.

Figure 1a shows scanning electron microscope (SEM) images of a fabricated device. The active patch-cavity area is 50 × 50 µm², with a square patch side L=1.45 µm, which sets the photonic resonance, and a period P=3L. The patches top surfaces are electrically connected by a 150-nm-thick Au wire, and coupled to the signal line of a coplanar waveguide through Au tapered air bridges. These modulators exhibit a −3 dB cutoff frequency of 16.7 GHz and a broadband spectral response spanning 200 cm⁻¹ around 1150 cm⁻¹ (8.7 µm), as shown in Fig. 1b. The modulation contrast (defined as the difference between the reflectivity measured at positive and negative voltages, normalized by the larger of the two values, $|R_{+V} - R_{-V}|/\max(R_{+V},R_{-V})$) is extracted from FTIR reflectivity spectra acquired at different DC bias voltages. It reaches up to 80% at ±7 V and reveals the two polariton branches at 1110 cm⁻¹ and 1175 cm⁻¹. The RF spectral response measured using a tunable CW laser, while driving the modulator with a 15 dBm sinusoidal signal at 10 MHz, confirms broadband operation. Maximum modulation is obtained at the upper polariton wavelength of 1175 cm⁻¹ (see also Supplementary Note I). Overall, this new device generation improves all the key figures of merit compared to the previous generation [27]: increased modulation speed (from 8.6 to 16.5 GHz), broader spectral bandwidth, and higher modulation contrast (from 25% to 80%).

It is established that driving an intensity modulator with a train of short electrical pulses—equivalent to an electrical frequency comb in the RF domain—can transfer a CW laser line into an EO FC centered around the optical carrier [6], if the modulator is efficient enough. The high performance of our modulator enables such operating mode, when it is driven with a train of short (~100 ps) electrical pulses. The details of the comb generation will be discussed in the following sections. Here, we explain the operating principle of the EO comb analysis. In particular, we show that it is possible to obtain spectral measurements with a single FC directly using an electrical spectrum analyzer (ESA), without any optical interferometer. The schematic is shown in Fig. 1c for the cases with and without an absorber. Several EO

FC lines are generated symmetrically around the laser carrier frequency (we employ an external-cavity tunable laser from Daylight Solutions). When the EO comb is sent to a fast quantum-well infrared photodetector (QWIP) [29], heterodyne beat notes between all pairs of comb lines are generated. However, since the carrier laser is more intense than the sidebands, the dominant beat note occurs between each comb line and the carrier. This allows the optical comb to be directly down-converted to the RF domain on an ESA connected to the QWIP output, without the need for an optical interferometer. Figure 1d illustrates the EO FC assuming absorption occurs on only one side of the carrier line. The strong carrier frequency is $f_c$, and the comb lines are spaced by the repetition rate $f_{rep}$ (MHz range), with the frequency of mode $j$ ($j = 0, \pm 1, \pm 2, \ldots$) given by $f_j = f_c + jf_{rep}$. The beating between the two optical comb lines $f_{\pm j}$ and the carrier frequency $f_c$ produces the same RF frequency $jf_{rep}$, resulting in a folding effect of the two sides (and a reduction of the apparent absorption depth in the RF spectrum) (Fig. 1e). Despite this folding effect, the approach offers significant advantages, including the simplified experimental setup compared with traditional optical interferometers.

### 2.2. Single- and dual-frequency-comb generation

We have established that we can transfer a CW laser line into an EO FC centered around the optical carrier [6]. A key advantage of this approach is the tunability of the pulse train repetition rate, which directly determines the comb mode spacing in the frequency domain. A complete schematic of the experimental setup used in this work is shown in Fig. 2a. A tunable mid-IR laser is routed through polarization optics and beam splitters to the EO intensity modulator. The reflected signal is collected and detected using a fast QWIP [29], then amplified and analyzed with an ESA. Figure 2b shows the time-domain electrical pulse measured at the output of one pulse generator: it has a full width at half maximum (FWHM) of ~115 ps. Using a single pulse generator and a 20 mW CW laser tuned to 1175 cm$^{-1}$, where the modulator exhibits its highest responsivity, we successfully generate EO frequency combs with RF spectra extending up to 20 GHz for pulse repetition rates of 1 GHz, 500 MHz, 100 MHz, and 10 MHz (the lower limit of our pulse generator), as shown in Fig. 2c. The discrete and evenly spaced RF comb lines demonstrate tunable repetition rates spanning more than two orders of magnitude. The RF comb-line power exceeds the ESA noise floor by up to 70 dB at low frequencies, 50 dB at 10 GHz, and remains on the order of 20 dB at 20 GHz. We note that the RF spectra are not corrected for the frequency response of the QWIP (see Supplementary Note II).

Building on the generation of a single EO frequency comb, the natural next step is to extend this approach to dual-comb operation. Conventionally, dual-comb generation relies on mixing two independent EO combs produced by two intensity modulators illuminated by two laser beams and driven by two pulse generators with slightly different repetition rates. Here, we propose an innovative scheme that requires only a single laser beam and a single intensity modulator, which is directly driven by two phase-locked pulse generators with slightly different repetition rates to produce a dual-EO FC.

As illustrated in Fig. 3a, the two pulse trains, with repetition rates ($f_1$ = 40 and $f_2$ = 40.08 MHz), are combined using an RF power combiner and directly analyzed on an ESA. The resulting dual electrical comb is shown in Fig. 3b, with close-up views at 1 GHz, 3 GHz, and 5 GHz (Fig. 3c), where the frequency offset between the two combs increases linearly with frequency, as expected from the difference in repetition rates. The dual electrical comb was then applied to the EO intensity modulator, as shown schematically in Fig. 3d. The reflected signal from a CW laser (20 mW, 1175 $cm^{-1}$) was detected using the fast QWIP, amplified, and analyzed with the ESA. The resulting dual optical frequency comb is presented in Fig. 3e, with zoomed-in regions in Fig. 3f demonstrating that the frequency offset between the two optical combs closely follows that of the electrical comb, confirming the faithful transfer of the dual-comb structure from the RF to the optical domain.

Finally, to further confirm the dual optical comb without fast detection, the QWIP was replaced by a commercial VIGO detector (see Fig. 2a) with a bandwidth of approximately 800 MHz, followed by a 100 MHz low-pass filter (LPF) (Fig. 3g). In this configuration, we can use the well-known mapping formula from the optical to the RF domain, $\Delta\nu = f_r^2/(2\Delta f)$, where $\Delta\nu$ is the maximum optical frequency span that can be mapped without aliasing, $f_r$ is the FC repetition rate, and $\Delta f$ is the repetition-rate difference between the two FCs. This condition ensures that the down-converted RF comb remains within a span of $f_r/2$, thus avoiding spectral overlap.

With our parameters, a 10 GHz-wide optical comb is down-converted to a 20 MHz RF span with a mode spacing equal to the repetition-rate difference ($\Delta f$ = 0.08 MHz), as shown in Fig. 3h. Note: owing to parasitic low-frequency signals, the measurement was intentionally performed in a higher folding region of the down-converted spectrum, corresponding to beating between modes $jf_2$ and $(j\text{-}1)f_1$, observed between 40 and 60 MHz. The RF comb-line power in the MHz range is lower than that of the GHz range because, in the MHz down-converted regime, the signal arises from beating between the sidebands themselves rather than from beating with the strong optical carrier; nevertheless, the RF comb lines remain 30–40 dB above the noise floor. Multiple folding regions of the down-converted comb are shown in Fig. 3i.

### 2.3. Frequency-comb spectroscopy demonstration

Because the mid-IR spectral region contains strong molecular fingerprint signatures, the ability to generate both single- and dual-EO FC in the mid-IR using a single intensity modulator and a single laser enables spectroscopic applications on a compact platform. As a first proof of principle, we performed transmission measurements using a 10 cm-long, double-polished Ge etalon. The etalon provides a well-defined and periodic spectral response with a nominal FSR of 500 MHz (0.0167 $cm^{-1}$), making it an ideal reference sample to validate the spectral resolution capability of the system. Our platform allows flexible operation between single- and dual-comb configurations. Direct detection is possible operating in the GHz range (using fast QWIP). Alternatively, if we choose to down-convert, we can operate in the MHz range with the standard dual-comb approach (using a Vigo detector and a 100 MHz LPF). Detection in the GHz range, i.e. single-comb detection, provides higher RF comb-line power through beating with the stronger optical carrier. In contrast, dual-comb (MHz-range) detection simplifies the setup by eliminating the need for a fast QWIP, at the cost of reduced RF power per comb line.

We first demonstrate transmission spectroscopy with direct detection in the GHz. To increase the number of available comb lines within the measurement bandwidth, we use a dual-comb. It is equivalent to a "single" comb with an increased number of comb lines within the measurement bandwidth (Fig. 4a).

When the dual EO comb (20 MHz and 20.02 MHz repetition rates, 20 mW incident CW laser power at 1175 $cm^{-1}$) is transmitted through the Ge etalon, the periodic Fabry–Perot resonances modulate the optical comb envelope. After detection by the fast QWIP, these modulations appear as periodic amplitude variations across the RF comb lines. The sample-induced transmission spectrum is obtained by subtracting, in dB (equivalent to a normalization with a reference), the RF spectra recorded with and without the sample (Fig. 4b). The resulting power differences between corresponding RF comb lines clearly reveal Fabry–Pérot oscillations with a FSR consistent with the etalon length and refractive index, measured to be approximately 500 MHz. This result confirms that the optical spectral response of the sample is faithfully mapped into the RF domain and can be directly analyzed using an ESA.

Following this validation, we demonstrate molecular absorption spectroscopy using an $NH_3$ gas cell. The Ge etalon is replaced by a glass cell containing $NH_3$ at a pressure of 10 mbar to limit collisional broadening. Transmission spectra are recorded using the same detection principle (Fig. 4c). As in the etalon measurement, the transmission spectrum is retrieved from the power ratio (difference in dB) between RF comb lines measured with and without the gas cell. Figure 4d shows the 10 GHz RF comb spectra and the corresponding transmission spectra obtained by tuning the external-cavity CW laser to wavenumbers of 1176.7 $cm^{-1}$ and 1176.9 $cm^{-1}$, respectively. These operating frequencies are chosen to stitch together two regions of interest and probe all the $NH_3$ absorption lines within a 20 GHz window around 1177 $cm^{-1}$, as referenced by the HITRAN database (Fig. 4e), where the modulator exhibits its highest responsivity. At the $NH_3$ pressure considered, the FWHM of the absorption linewidths is on the order of 200–300 MHz: it well matches the spectral sampling provided by the EO comb. Owing to the limited absolute frequency accuracy of our external-cavity laser (Daylight Solutions MirCAT1000), the nominal and actual wavenumbers may differ slightly. The spectra are therefore aligned using the molecular absorption lines from the HITRAN database as an absolute frequency reference (see Supplementary Note III). As a proof of principle, the analysis is limited to a 10 GHz RF span to maintain a sufficiently high signal-to-noise ratio. The measured absorption features show excellent agreement with the HITRAN reference spectra over a 20 GHz optical bandwidth, and all the observed lines can be unambiguously assigned to known $NH_3$ transitions. These results demonstrate both the spectral accuracy and sensitivity of the system. They prove that robust spectroscopic performance is achieved despite the spectral folding inherent to RF-domain detection on the ESA.

After using a fast QWIP and a "single" comb configuration, we exploit the approach conventionally referred to as DCS (dual comb spectroscopy). It relies on a slow detector to measure a down-converted dual (EO) comb in the MHz range, generated through the heterodyne beating of two combs with slightly different $f_{rep}$. In this configuration, a dual EO comb with repetition rates of $f_1'$ = 40 MHz and $f_2'$ = 40.08 MHz (20 mW incident CW laser at 1175 $cm^{-1}$) is transmitted through a Ge etalon and detected using a VIGO detector equipped with a 100 MHz LPF to mimic a slow detector. The resulting signal is analyzed by an ESA (Fig. 5a,c). As discussed previously, owing to parasitic low-frequency signals, the measurement is performed by analyzing the beating between modes $jf_2'$ and $(j\text{-}1)f_1'$, rather than between $jf_2'$ and $jf_1'$, corresponding to a RF window between 40 and 60 MHz. The effective mode spacing between $jf_2'$ and $(j\text{-}$

*1)f₁′* remains 40 MHz, which is more than ten times smaller than the FSR of the Ge etalon. Figure 5c shows the RF spectra of the down-converted dual comb measured with and without the Ge etalon. The corresponding transmission spectrum, obtained from the power differences between matched RF comb lines, clearly reveals Fabry–Pérot oscillations with a FSR of approximately 500 MHz. To further validate the characterization of the Ge etalon, conventional transmission measurements were also performed using an 8.5 µm DFB-QCL (Thorlabs), whose wavelength was finely scanned via a low-noise current driver and detected using a liquid-nitrogen-cooled MCT detector with lock-in amplification (Fig. 5b,d). The measured FSR of ~500 MHz agrees well with the Fabry–Pérot resonances retrieved from both the "single" comb and down-converted dual-comb measurements, confirming the accuracy and robustness of the EO comb spectroscopy approach.

## 3. Conclusions

In summary, we have demonstrated a compact platform for generating mid-IR single- and dual-EO FCs around 9 µm (but the same approach is in principle applicable across the full mid-IR spectral range). This work provides three key advances: (i) ultrafast, broadband mid-IR amplitude modulators with very high modulation contrast; (ii) the use of a single modulator and a single CW laser to produce both single- and dual-EO combs by driving the modulator with one or two phase-locked electrical pulse trains; (iii) the direct observation of the optical comb structure in the RF domain using an ESA, enabled by the strong laser carrier serving as the main beating line for spectroscopy applications.

The EO FCs are demonstrated up to 20 GHz around the tunable carrier laser, with mode spacings adjustable from 10 MHz (limited by the pulse generator, not by the modulator) to GHz. As a proof of principle, broadband spectroscopy was performed on a Ge etalon (FSR ~500 MHz) and a 10 mbar $NH_3$ gas cell (linewidth ~300 MHz), successfully resolving Fabry–Pérot resonances and molecular absorption lines with good agreement with HITRAN reference data. Both fast and slow detection schemes, corresponding to "single" and dual-comb spectroscopy, confirm high spectral fidelity and resolution.

This platform provides broad tunability and an interferometer-free approach to mid-IR spectroscopy, opening pathways for practical FC systems for precision spectroscopy in the mid-IR molecular fingerprint region.

## 4. Method

### 4.1. Device fabrication

The fabrication process requires five lithography steps. First, the epitaxially grown wafer and a semi-insulating GaAs host wafer are coated with 10/300 nm Ti/Au and bonded together by thermal-compression wafer bonding. The InP substrate is then removed using HCl to expose the active region. The top metal patches, electrically interconnected by thin Au wires, are subsequently defined using electron-beam lithography (EBL), followed by the deposition of 10/250 nm Ti/Au and lift-off. An inductively coupled plasma (ICP) etching step is then performed to etch the active region, using the top metal patches as an etch mask. Next, the coplanar waveguide (CPW) is patterned using optical mask lithography, followed by 10/200 nm Ti/Au deposition and lift-off. A photoresist scaffold covering the gap between the CPW hot electrode and the top metal is patterned by optical lithography and reflowed at elevated temperature to obtain a smooth, sloped profile. Finally, a second EBL step is used to define tapered Au bridges on top of the scaffold, connecting the top metal patches to the CPW hot electrode and enabling electrical biasing of the device. The device is then wire-bonded to an external CPW terminated by a 2.92 mm (K-type) solderless connector.

### 4.2. Experimental setup

The complete RF setup used in this work is shown in Fig. 2a. For the optical bandwidth characterization of the modulator (Fig. 1b), a sinusoidal signal from an RF signal generator is fed to the modulator, and the reflected signal from a 15 mW CW laser at 1175 $cm^{-1}$ is sent to a QWIP. The QWIP detects the beating between the carrier and generated sidebands, producing a beat note at the frequency set by the synthesizer. This beat note is measured and analyzed using an ESA. The RF source power is adjusted to compensate for cable losses between the RF source and the modulator, ensuring a consistent dBm level throughout the frequency scan. The measured RF response is corrected for the QWIP frequency response and divided by a factor of 2, since the ESA displays electrical power in dBm (proportional to voltage squared), whereas the optical power is already proportional to the voltage generated by the QWIP.

For the EO comb RF spectra reported in this work, all the measurements are directly recorded on the ESA without any post-correction. In these cases, the QWIP signal is amplified by a 30 dB RF amplifier.

## Supplementary information

Supplementary information is available for this paper.

## Acknowledgements

This work was carried out within the C2N micro- and nanotechnology platforms and was partly supported by the RENATECH network and the General Council of Essonne. We thank G. Aubin for help on RF measurements, D. Morini and V. Turpaud for the loan of the pulse generator, and E. Peytavit for help with detector fabrication. We acknowledge financial support from the ERC Advanced "SMART-QDEV" (Grant No. 101095959), the ANR "MIR-SESAM" (Grant. No. ANR-23-CE24-0014), and the Programme de Prematuration of CNRS.

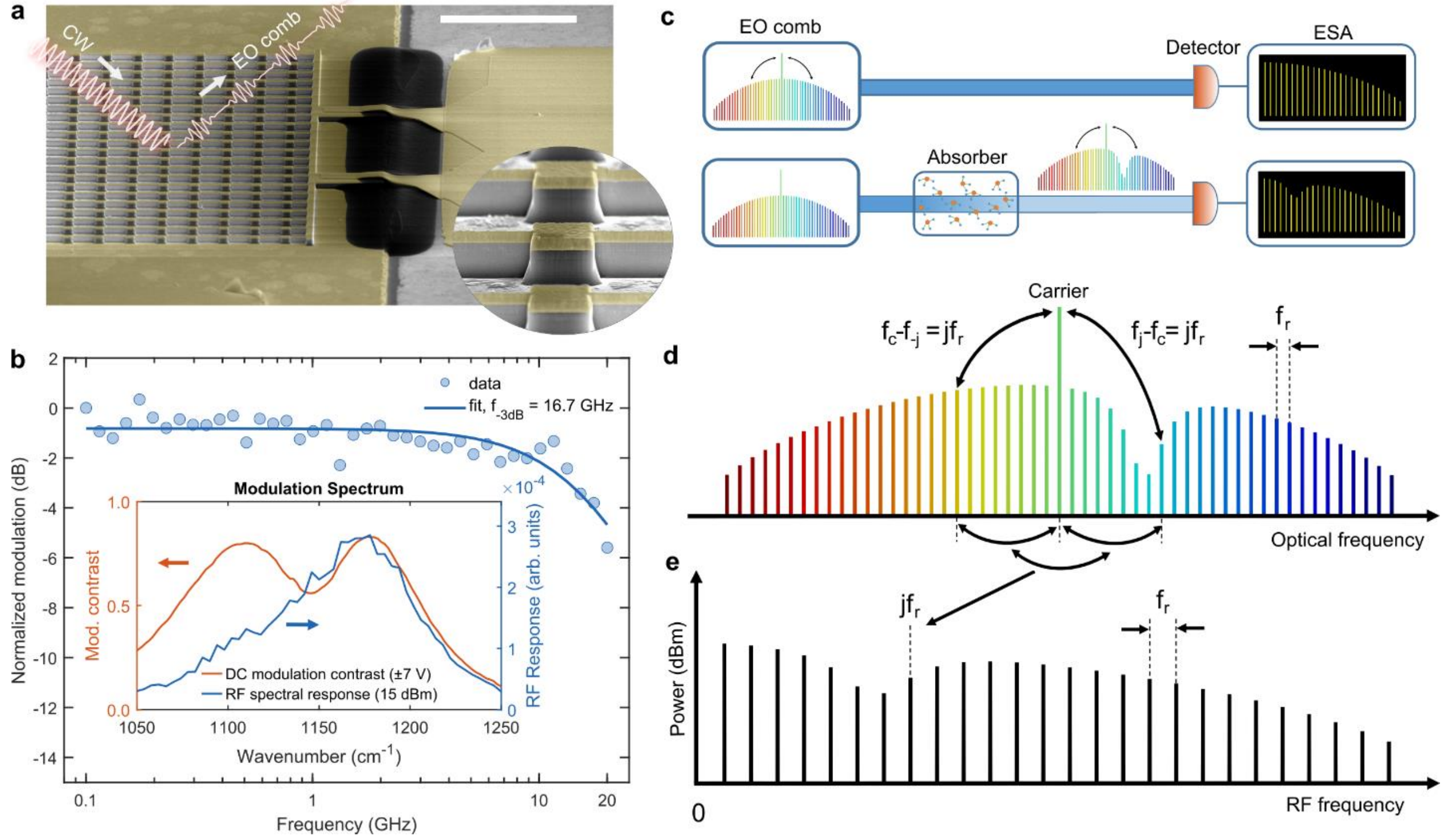

**Fig. 1 | Mid-IR intensity modulator characteristics and operating principle of direct analysis of an EO FC using an electrical spectrum analyzer (ESA). a** SEM image of a typical electro-optic modulator, showing the patch area used for CW optical injection and EO comb generation in reflection. Yellow false coloring represents the Ti/Au contacts. The white scale bar corresponds to 20 µm. The inset is a close-up on a single patch with QWs embedded in the metal–metal cavity.

**b** Normalized modulation response of the modulator as a function of RF frequency. Dots represent measured data, and the solid line is a fitted curve (first-order low-pass filter), yielding a 3-dB bandwidth of 16.7 GHz. Inset: modulation spectrum showing the DC modulation contrast (orange) and RF spectral response (blue) versus laser wavenumber. The RF spectral response was measured at a modulation frequency of 10 MHz. See also methods and Supplementary Note I.

**c** Schematic of the experimental measurement setup. The generated EO comb is directly detected by a fast photodetector and analyzed using an ESA (top). In the lower path, an absorber is introduced to modify the optical spectrum before detection, enabling comparative spectral analysis.

**d** Illustration of the optical frequency-domain structure of the EO comb, assuming that absorption appears on only one side of the carrier laser line. The strong carrier frequency $f_c$ and comb lines spaced by the repetition rate $f_r$ are shown. The frequency of mode j (j = 0, ±1, ±2, …) is given by $f_j = f_c + jf_r$.

**e** Corresponding RF spectrum obtained on the ESA, showing discrete RF comb lines at integer multiples of the repetition frequency $f_r$. The RF comb line at $jf_r$ arises from the beat note between the two optical comb lines $f_{\pm j}$ and the carrier laser frequency $f_c$.

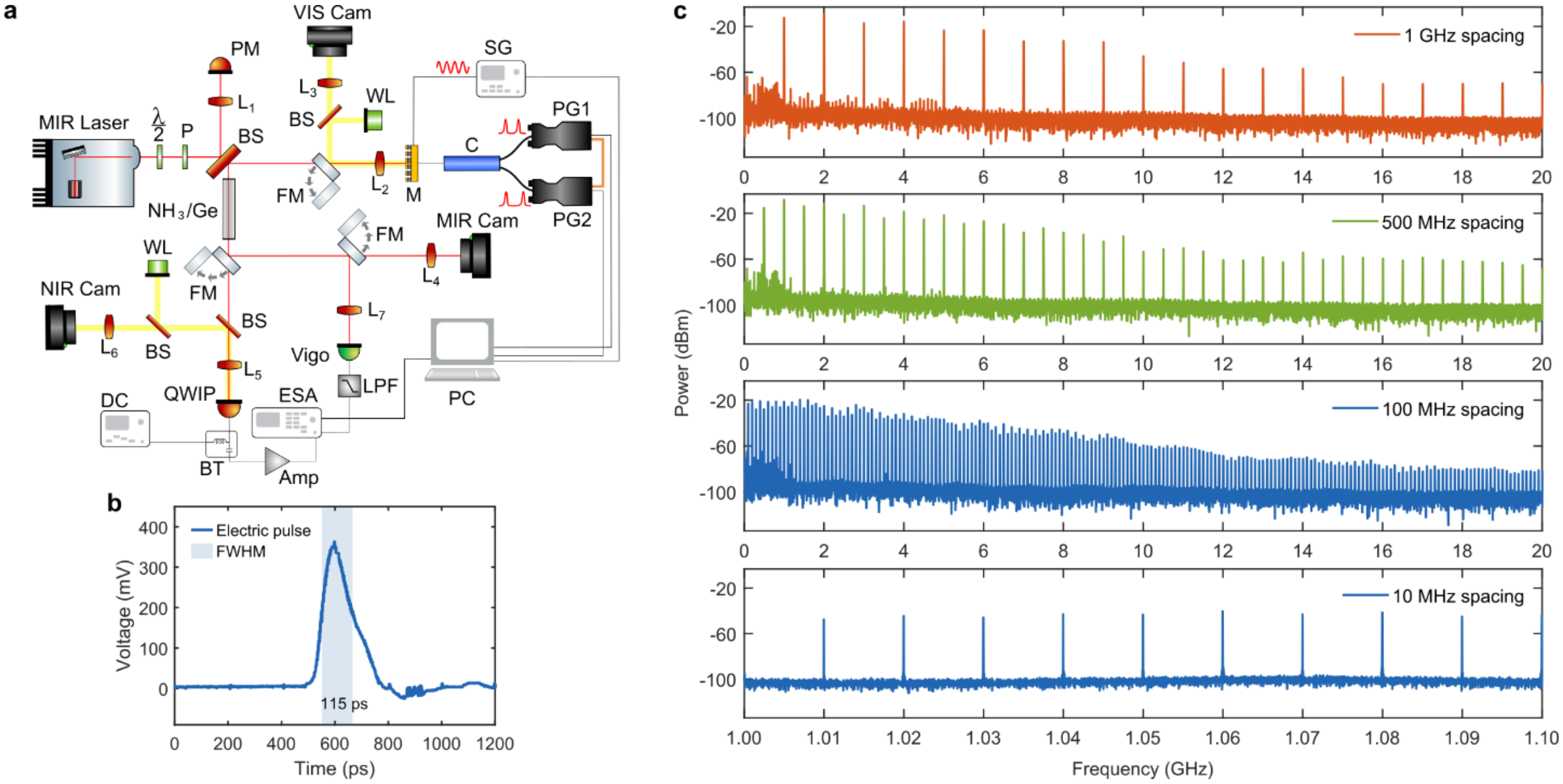

**Fig. 2 | Experimental setup and RF spectral characteristics of single-EO FC generation. a** Schematic of the full experimental setup used in this work. WL: white-light source; P: polarizer; λ/2: half-wave plate; BS: beam splitter; PM: power meter; SG: RF signal generator; PG: pulse generator; C: RF power combiner; L: lens; DC: Keithley source meter; $NH_3$/Ge: $NH_3$ gas cell or Ge etalon, M: modulator, Cam: camera, FM: flipped mirror; Vigo: MCT detector; LPF: low-pass filter; ESA: electrical spectrum analyzer; BT: bias tee; PC: computer; Amp: RF amplifier. **b** Time-domain electrical pulse measured at the output of one PG, showing a FWHM of approximately 115 ps. **c** RF spectra of the EO comb measured on the ESA for different pulse repetition rates, showing comb line spacings of 1 GHz, 500 MHz, 100 MHz, and 10 MHz (limited by the PG). Discrete and evenly spaced RF comb lines demonstrate tunability of the repetition rate over more than two orders of magnitude. Only a single PG is used in this measurement, and no absorber ($NH_3$/Ge) is inserted.

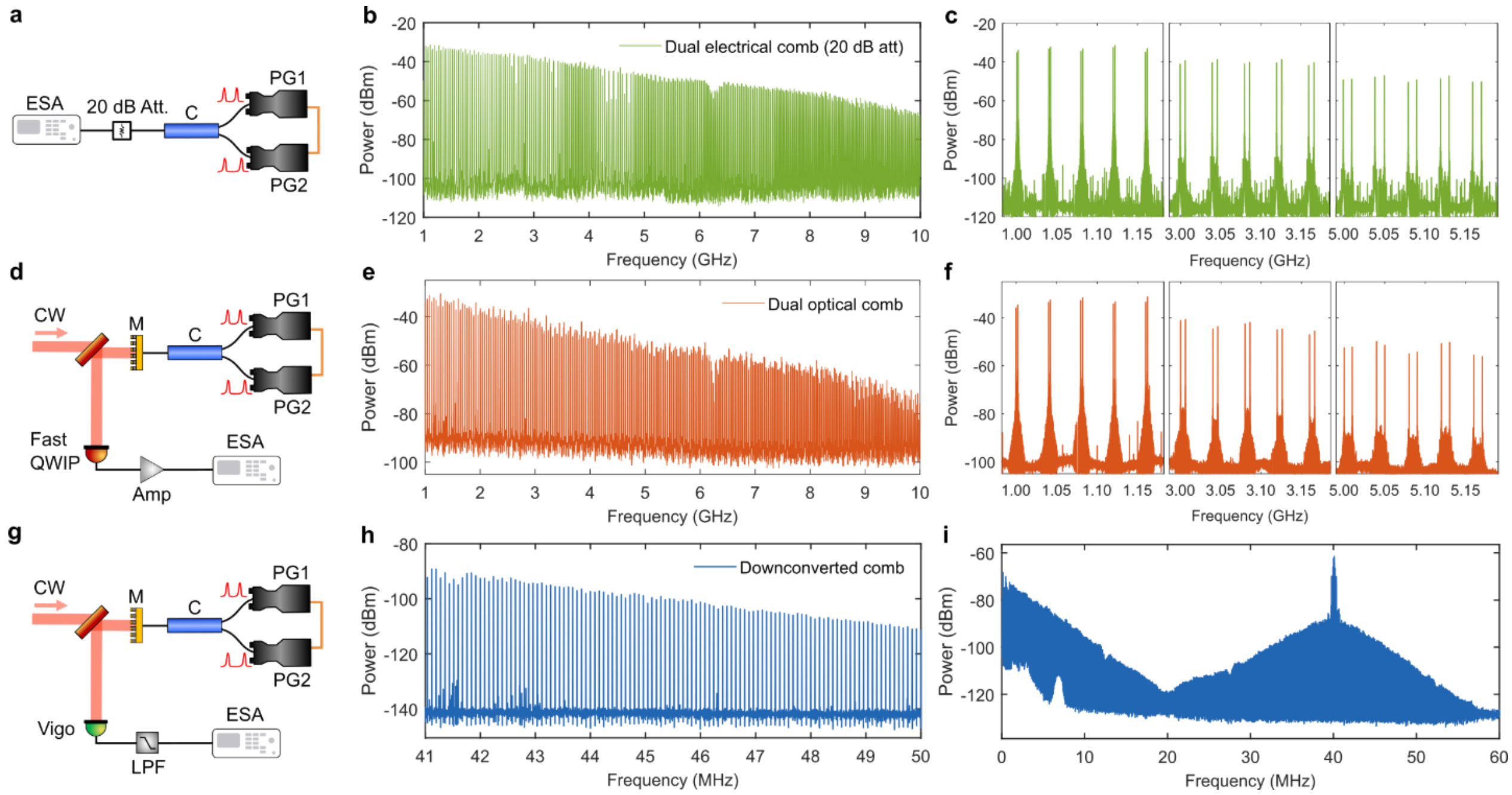

**Fig. 3 | RF spectral characteristics of dual-EO FC generation. a** Schematic of the measurement presented in **b** and **c**. Two pulse generators with slightly different repetition rates ($f_1$ = 40 MHz and $f_2$ = 40.08 MHz) are fed into a RF power combiner and sent to the ESA with 20 dB attenuation. **b** RF spectrum of the dual electrical FC. **c** Zoomed-in regions around 1 GHz, 3 GHz, and 5 GHz, showing that the frequency offset between the two combs increases with frequency. **d** Schematic of the measurement presented in **e** and **f**. The dual electrical comb shown in **b** is applied to the intensity modulator. The reflected signal from a CW laser is detected by a fast QWIP, amplified by an external amplifier, and analyzed using the ESA. **e** RF spectrum of the dual optical FC. **f** Zoomed-in regions around 1 GHz, 3 GHz, and 5 GHz, showing that the offset between the two optical combs increases with frequency and closely follows the electrical comb shown in **c**. **g** Schematic of the measurement presented in **h** and **i**. Similar to **d**, but the fast QWIP is replaced by a commercial VIGO detector with a bandwidth of ~800 MHz and filtered with a 100 MHz LPF to mimic a slow detector. **h** RF spectrum of the down-converted optical FC with a comb spacing equal to the difference between the repetition rates of the two pulse generators $\Delta f = f_2 - f_1$ = 0.08 MHz. A 10 GHz optical comb bandwidth is down-converted to 20 MHz. Due to parasitic low-frequency signals, the measurement is intentionally performed at a higher folding region of the down-converted comb, corresponding to the beating between modes $jf_2$ and $(j\text{-}1)f_1$, observed between 40 and 60 MHz. **i** RF spectrum of the down-converted optical frequency comb showing multiple folding regions.

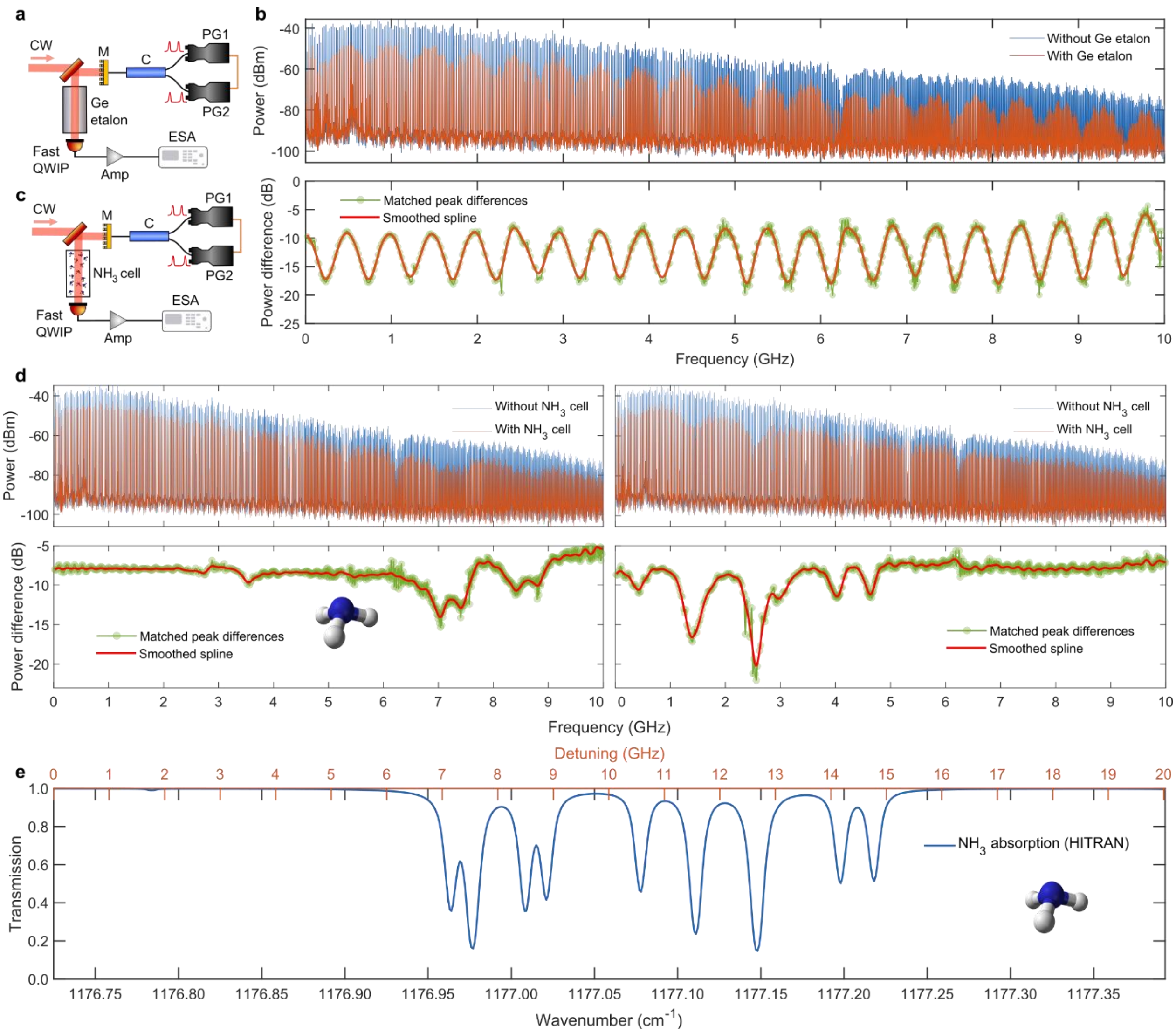

**Fig. 4 | Frequency-comb spectroscopy (FCS) using a fast QWIP and an ESA. a** Schematic of the measurement shown in **b**. The dual-EO FC (20 MHz and 20.02 MHz) is transmitted through a 10 cm-long Ge etalon with a FSR of ~500 MHz, then detected by a fast QWIP and analyzed using an ESA. This experiment can equivalently be performed with a single comb, that is, using only one PG. **b** (Top panel) RF spectra of the dual EO comb without (blue) and with (red) the Ge etalon. (Bottom panel) Power differences between matched comb lines, showing clear Fabry–Pérot oscillations with a FSR of ~500 MHz. The red curve is a smoothing spline of the raw data (transparent green dots). **c** Schematic of the measurement shown in **d**, similar to **a**, but with the Ge etalon replaced by a 10 cm-long $NH_3$ gas cell at a pressure of 10 mbar. **d** Same analysis as in **b**. The top panel shows RF spectra of the dual EO comb with and without the $NH_3$ cell, while the bottom panels show power differences between matched comb lines. The left half corresponds to the tunable external-cavity CW laser nominally set to a wavenumber of 1176.7 $cm^{-1}$, and the right half corresponds to 1176.9 $cm^{-1}$. **e** Reference $NH_3$ absorption lines from the HITRAN database over a 20 GHz optical bandwidth centered around 1177 $cm^{-1}$. All absorption features detected in **d** match well with the HITRAN database. Owing to the limited absolute accuracy of the external-cavity laser, the nominal and actual wavenumbers may differ slightly. We therefore used the measured absorption lines as an absolute frequency reference.

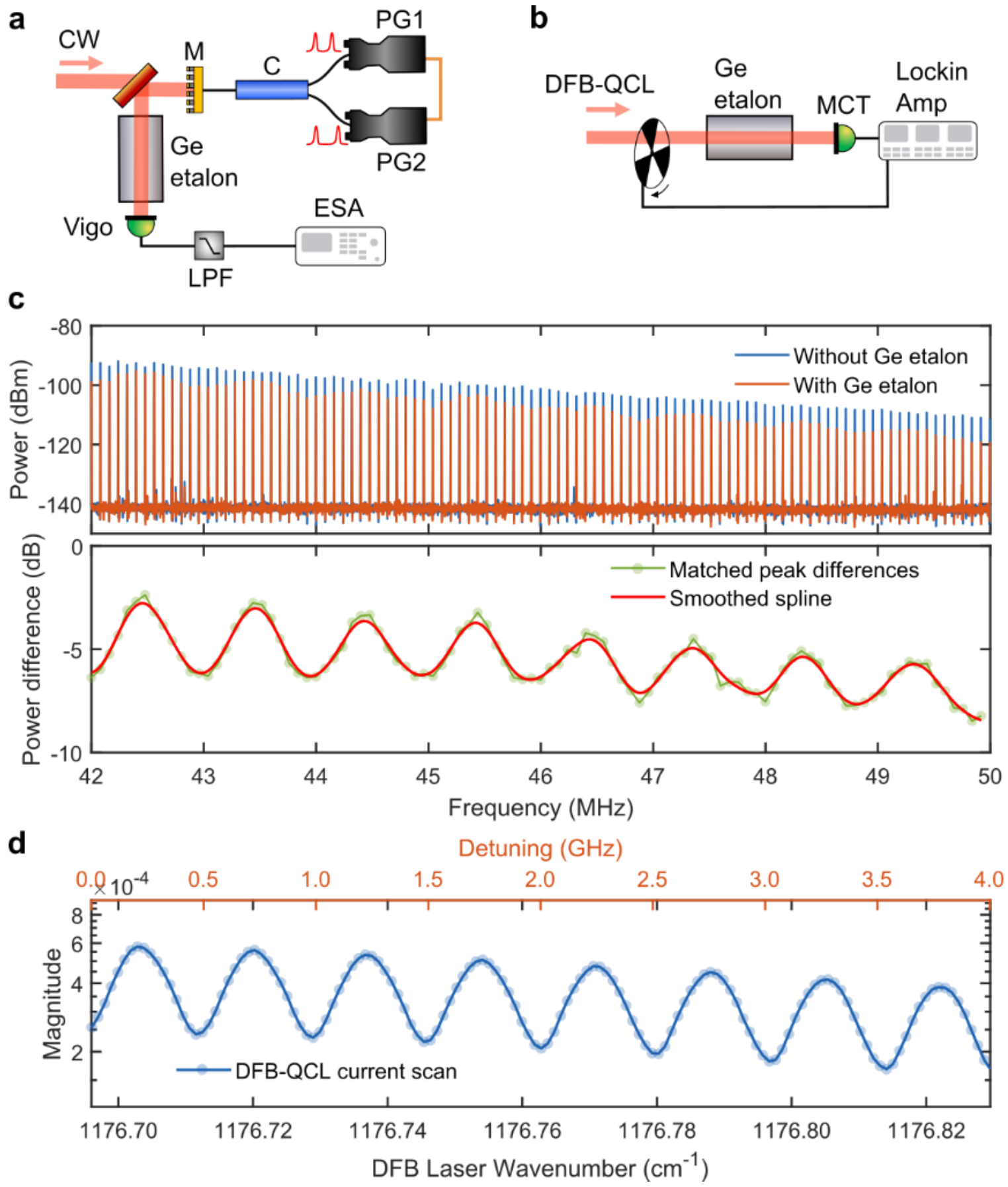

**Fig. 5 | DCS using a slow detector and an ESA. a** Schematic of the measurement shown in **c**. A dual-EO comb (40 MHz and 40.08 MHz) is sent through a Ge etalon and downconverted on a VIGO detector with a 100 MHz LPF, then analyzed using an ESA. **b** Schematic of the measurement shown in **d**. A DFB-QCL is sent through an optical chopper and the Ge etalon, then detected by a liquid-nitrogen-cooled MCT detector with a lock-in amplifier. The current of the DFB laser is finely tuned using a low-noise current driver to sweep the optical wavenumber. **c** (Top panel) RF spectra of the downconverted EO comb without (blue) and with (red) the Ge etalon. (Bottom panel) Power differences between matched comb lines, showing clear Fabry–Perot oscillations with a FSR of ~500 MHz. **d** Transmission measurement of the Ge etalon using the DFB laser, showing an FSR of ~500 MHz, which confirms the FCS measurement.